# Chiral ferroelectric nematic liquid crystals as materials for versatile laser devices

César L. Folcia[a], Josu Ortega[a], Teresa Sierra[b], Alejandro Martínez-Bueno[b], and Jesús Etxebarria[a]

[a]Department of Physics, Faculty of Science and Technology, University of the Basque Country, UPV/EHU, Bilbao, Spain
[b]Instituto de Nanociencia y Materiales de Aragón (INMA), Departamento de Química Orgánica, Facultad de Ciencias, CSIC-Universidad de Zaragoza, 50009 Zaragoza, Spain

Email: cesar.folcia@ehu.es or josu.ortega@ehu.es

We present a liquid-crystal laser device based on the chiral ferroelectric nematic phase ($N_F^*$). The laser medium is obtained by mixing a ferroelectric nematic material with a chiral agent and a small proportion of a fluorescent dye. Notably, in the $N_F^*$ phase very low electric fields perpendicular to the helical axis are able to reorient the molecules, giving rise to a periodic structure whose director profile is not single harmonic but contains the contribution of various Fourier components. This feature induces the appearance of several photonic bandgaps whose spectral ranges depend on the field, which can be exploited to build tunable laser devices. Here we report the characterization of home-made $N_F^*$ lasers that can be tunable under low electric fields and present laser action in two of the photonic bands of the material. The obtained results open a promising route for the design of new and more versatile liquid-crystal based lasers.

**Introduction**

Photonic materials are periodic dielectric structures which show wavelength ranges where the propagation of some electromagnetic waves is forbidden. These spectral regions are called photonic band gaps (PBGs), and at their edges, the density of optical states shows a sharp increase. The high values of the density of states can be exploited for many applications including light amplification, leading to the so-called distributed feedback (DFB) laser action. Cholesteric liquid crystals (CLCs) are the most prominent examples of photonic materials among mesogenic compounds. They are constituted by chiral molecules that self-assemble forming a helical structure with a certain helical pitch. In these materials, it can be shown that a PBG results for circularly polarized light with the same handedness as that of the helix, which cannot propagate along the helix axis for a spectral region between $\lambda_o = pn_o$ and $\lambda_e = pn_e$, being $p$ the helical pitch, and $n_o$ and $n_e$ the ordinary and extraordinary local refractive indices respectively. Hence, a mirrorless DFB type laser emission can be achieved at the short-wavelength edge (SWE) or long wavelength edge (LWE) of the gap when CLCs are doped with fluorescent dyes whose emission spectra overlap with the PBG [1–5]. Since the discovery of laser emission in CLCs by Kopp et al. in 1998 [1], many efforts have been made to build up CLC lasers with increasing performance, opening an extensive research field. This type of lasers presents useful characteristics, such as low pumping power threshold and ease of implementation in small devices. On the other hand, one of the most attractive features of photonic CLC lasers is the possibility of tuning the PBG by external stimuli, with the consequent shift of the emission wavelength. Examples of different external agents for laser tuning are light irradiation [6–9], temperature variation [10,11] or mechanical stretching [12].

Especially interesting is the electric-field-induced tuning [13–17], since electric fields can be easily applied to the liquid crystal materials within the laser devices. Some examples of LC lasers tunable by electric fields are based on the oblique heliconical CLC (CLC$_{OH}$) state that certain CLCs display [18,19]. In these materials, the tuning can be driven by the strength or by the frequency of the electric field. Although the tuning range of the PBG is very wide in both cases, the molecular director makes an acute angle with the helix axis. This fact gives rise to a small contrast of the dielectric modulation of the structure, which produces an important decline of the density of states at the edges of the PBG and consequently a poor performance of the laser. A different approach of tunable lasers under applied field uses complex cells with polymer CLC mirrors whose pitch can be changed [20].

In the present work, we present a tuning alternative using a laser device based on the ferroelectric nematic ($N_F$) phase [21–30] recently discovered. Usually, this phase becomes stable when cooling the ordinary nematic phase N, although it can also appear directly from the isotropic liquid [31]. In the $N_F$ phase the head-to-tail molecular invariance, characteristic of traditional LCs, is broken, giving rise to the appearance of a high macroscopic polarization (of the order of $\mu C/cm^2$) along the molecular director. Giant dielectric constants (of the order of $10^4$) [32] and a high non-linear optical susceptibility (of the order of 10 pm/V) [33] are also remarkable features of these materials. Moreover, a chiral version of the $N_F$ phase, the so-called chiral ferroelectric nematic phase ($N_F^*$), can be obtained by adding a small proportion of chiral component to the $N_F$ materials [34–36] or by slight chiral modifications in the chemical structure of those prototype molecules [37, 38].

The linear optical properties of the $N_F^*$ phase are the same as those of an N* (i.e. conventional CLC) structure. This is true even though the actual periodicity is equal to the pitch $p$ for the $N_F^*$ structure and to $p/2$ for CLCs, due to the head-to-tail invariance of the molecular director in the latter case. Therefore, the $N_F^*$ phase (as well as the CLC) shows a single selective reflection band at normal incidence centered at $\lambda = pn$, being $\lambda$ the vacuum wavelength of light and $n = \sqrt{(n_e^2 + n_o^2)/2}$ the mean refractive index, in the usual way. The periodicity $p$ of the $N_F^*$ structure becomes evident, however, when an electric field is applied perpendicular to the helical axis, because now the nematic director is reoriented according to the polarity of the field, favoring regions where the local polarization tends to be parallel to it, to the detriment of others where they are almost antiparallel [34–36]. As a consequence, the structure with periodicity $p$ cannot be described by a single-harmonic modulation but also contains distortions whose Fourier components have periodicities $p/m$ ($m = 2, 3, ...$). Therefore, a reflection band at $\lambda = 2pn$ appears together with bands centered at $\lambda_m = 2pn/m$ with reflectivity depending on the specific shape of the distortion and the relative magnitude of their Fourier contributions. On the other hand, the electric field induces an enlargement of the pitch in order to compensate the increase of the twist elastic energy in the regions where the molecular orientation changes rapidly [34,36]. This increase in $p$ gives rise to a redshift with the field in all the reflection bands. In summary, contrary to the case of the CLC$_{OH}$ where a single band centered at $\lambda_2 = pn$ appears [18,19], the reflectance spectrum of the N$_F$* structure under field shows various PBGs whose position, intensity, and shape depend on the applied field intensity.

Here, we report for the first time laser emission in a liquid crystal cell in the N$_F$* phase. The study covers laser emission characteristics, electric-field-induced tuning at the LWE of the $\lambda_2$ photonic band (the only one observed without field), and lasing at the third harmonic $\lambda_3$ band of the distorted helical structure.

**Experimental section**

RM734 was synthetized following the procedure described in [33]. S1 was previously described in [37] and was synthetized for this work following a modified procedure gathered in the supporting information.

The temperature of the samples during texture analysis, reflectance spectroscopy, and laser emission experiments was controlled with a hot stage INSTEC SC-200 with 0.1 °C accuracy. Home-made cells (18 − 20 μm thickness) with ITO strip electrodes separated 5 mm apart on one of the glasses for the application of in-plane electric fields were capillary-filled in the isotropic phase. Texture observation together with analysis of the electrooptic behaviour under AC voltages at 0.5 Hz frequency were performed with a polarizing microscope (Olympus BX51). Spectroscopic measurements in the N$_F$* phase were performed in the reflection mode with a fiber-optic spectrometer (Avantes – AvaSpecULS2048-USB2, grating: 300 lines/mm, slit size =100 μm) with $\Delta\lambda = 5$ nm of resolution full width at half maximum (FWHM), using light from a deuterium-halogen lamp.

For laser experiments, cells were optically pumped by using a Nd:YAG laser (LOTIS TII LS-2138T-100), operating at the second harmonic frequency (wavelength 532 nm). The laser pulse duration was 14 ns with repetition rates of 0.5 or 1 Hz for different experiments and its intensity was selected with a variable reflectance type attenuator (LOTIS II Attenuator 532 nm). Laser light was right (or left) circularly polarized with a variable circular polarizer for 532 nm (Thorlabs VC5-532/M) before being focused on the sample, at normal incidence, with a lens of 20 cm of focal length. Pumping pulse energy was measured with a power meter (Ophir), at a repetition rate of 20 Hz. Laser pulses and square-wave electric field applied to the material were synchronized and their relative delay was controlled. A 532 nm notch filter was placed behind the sample to remove the pumping laser light. Laser emission spectra were measured using a fiber-based spectrometer (AvaSpec 2048) with a resolution of 0.4 nm FWHM whose optical fiber was situated just behind the notch filter without any focusing lens.

**Experimental results and discussion**

The material employed in the experiments was a mixture of the $N_F$ liquid crystal 4-[(4-nitrophenoxy)carbonyl]phenyl2,4-dimethoxybenzoate (RM734) [21], the fluorescent dye 1,3,5,7,8-pentamethyl-2,6-di-t-butylpyrromethene-difluoroborate complex (PM597) (Sigma-Aldrich), and a chiral dopant to promote the helical structure. The type and proportion of the chiral dopant were selected in such a way that the PBG of interest was within the fluorescence range of the dye and, on the other hand, the local birefringence $\Delta n = n_e - n_o$ of the medium was reduced as little as possible, in order to preserve the contrast that defines the pitch periodicity. Samples were plane-parallel cells made of two glasses treated with polyvinyl alcohol and rubbed for parallel alignment. Thus, as in a conventional CLC laser, the material was aligned in the Cano geometry, i.e., with the helical axis perpendicular to the cell glasses. Typical cell thicknesses were 20 μm. On one of the glasses, two ITO strip electrodes separated 5 mm apart allowed for the application of in-plane electric fields, i.e., perpendicular to the helical axis. The material was introduced by capillarity into the cells in the isotropic phase. In all the cases the alignment was improved by applying low frequency (1 Hz) small AC voltages (~10 V) in the $N_F$* phase for several minutes.

First, the usual laser emission was checked and characterized at the LWE of the reflection band of the $N_F$* phase under zero field. We used a right-handed, chiral dopant: (13bS)-5,6-Dihydro-5-(trans-4-propylcyclohexyl)-4H-dinaphtho[2,1-f:1',2'-h][1,5]dioxonin

(R5011) (BLDpharm). RM734, R5011, PM597 were mixed in proportion 96.4:2.8:0.8 respectively (wt. %). With these concentrations the $N_F^*$ phase becomes stable at 120 ºC on cooling from the high temperature N* phase, and undergoes a transition to a crystalline solid phase at about 90 ºC. Figure 1(a) shows a typical texture of the sample in the $N_F^*$ phase as observed between crossed polarizers. Fig 1(b) represents the reflectance spectrum of the sample. In spite of the fact that the alignment was far from being ideal, laser emission was detected when the cell was optically pumped using a pulsed Nd:YAG laser operating at the second harmonic frequency (wavelength 532 nm, pulse duration 14 ns). The pumping light was left-handed circularly polarized, in order to optimize the excitation conditions, with a pulse energy of 3 µJ and a repetition rate of 1Hz. The light was focused on the cell at normal incidence and the spot was Gaussian, with a diameter $D4\sigma = 240$ µm (FWHM = 144µm). Figure 1(c) shows the laser emission observed when focusing on two different regions of the sample. A slight difference in the laser wavelength is detected due to small inhomogeneities in the pitch values across the sample area.

Black circles in Figure 1(d) represent the laser emission energy (in arbitrary units) for various pumping pulse energies. They show the typical dependence of a laser emission, i.e., a threshold energy from which the emission intensity grows almost linearly with the pumping intensity.

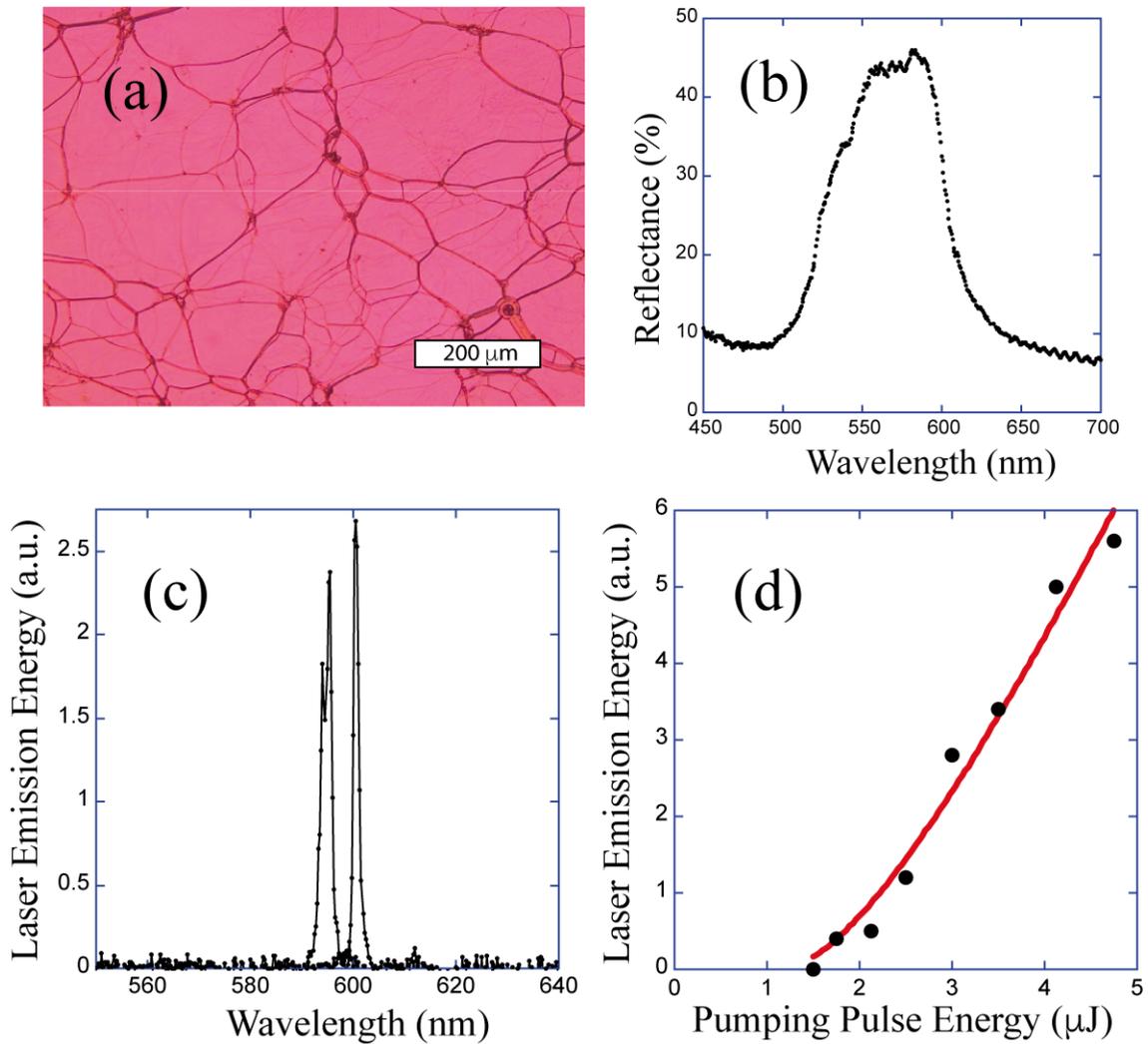

**Figure 1.** Optical characteristics and laser emission at the LWE of the reflection band under zero field. (a) Photomicrograph of the texture of the material at 108 °C in the $N_F^*$ phase as observed between crossed polarizers. (b) Reflectance spectrum of the sample at the same temperature. (c) Laser emission spectra at two different regions of the cell when the material was optically pumped with 3 µJ/pulse. (d) Laser emission energy (in arbitrary units) vs energy of the pumping pulse.

Taking into account the Gaussian profile of the illumination spot, it can be shown (5) that the energy emitted as laser radiation, $E_{laser}$, is related to the input energy $E_{in}$ according to the expression

$$E_{laser} = C\left[E_{in} - E_{th}\left(1 + \ln\frac{E_{in}}{E_{th}}\right)\right], \qquad (1)$$

where $E_{th}$ is the threshold energy and C is a constant related to the slope efficiency for $E_{in} \gg E_{th}$. From a fit of the experimental points in Fig. 1(d) to expression (1) (red line) the parameters $C = 2.9 \pm 0.3$ and $E_{th} = 1.1 \pm 0.1$ µJ/pulse result. It is interesting to point out that $E_{th}$ is comparable to typical threshold energies of conventional CLC lasers. The constant C is not relevant in the present study since $E_{laser}$ is expressed in arbitrary units. For example, $E_{th} = 1.2 \pm 0.1$ µJ/pulse was reported in Ref. 5 for a sample of 10 µm thickness with the same dye. Although $E_{th}$ should certainly increase somewhat for thinner samples, it seems that the performance of the laser based on the $N_F$* phase of the RM734 – R5011 mixture is quite acceptable despite its alignment imperfections.

Next, a square wave electric field of frequency 10 Hz was applied to the material. Figure 2 represents the selective reflection band corresponding to $\lambda_2 = pn$ for different field intensities. As a consequence of the pitch enlargement, the zero-voltage band is gradually (and reversibly) red shifted (up to about 20 nm) when the field is increased. For electric fields higher than about 11 V/mm the reflectance spectrum rapidly decreases and becomes less defined, indicating a significant distortion of the structure. Figure 3 shows laser emissions observed for different square-wave electric fields of frequency 1 Hz. The pumping laser was synchronized with the square voltage and was triggered just after the polarity inversion to reduce as much as possible the unavoidable shielding due to the migration of electrical charges from impurities. As in the spectra of Fig. 2, the electric field was varied monotonously and the effect was found to be reversible although, for relatively high fields, the material requires to remain at zero field for a certain time (about 1/2 minute) to recover the initial condition. Figure 3 shows also the fluorescence spectrum of the dye in the material (black dots). The laser intensity reduces as the dye fluorescence decreases. However, the tuning range is not only limited by the fluorescence range but also by the electric field intensity, i.e., lasing is hardly possible if the field produces a great structural distortion (see e.g. the PBG of Fig. 2 for 12.6 V/mm).

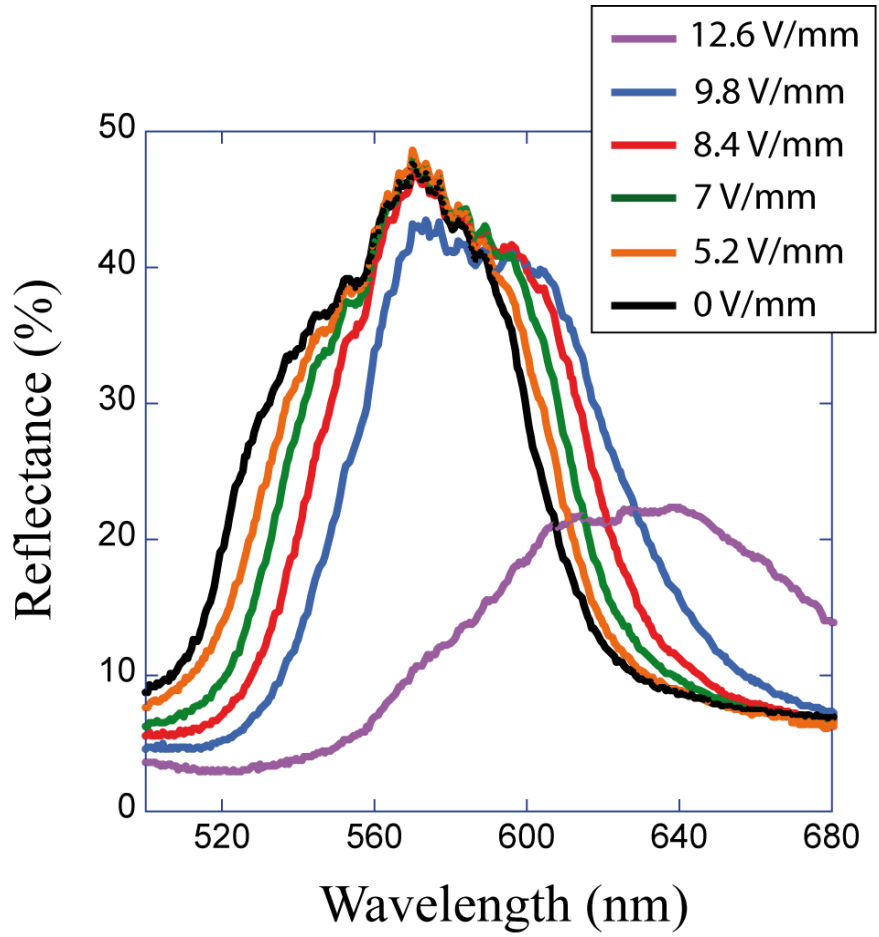

**Figure 2.** Time averaged reflectance spectra around the $\lambda_2$ photonic band measured for different square-wave electric field intensities of frequency 10 Hz. The structure becomes strongly distorted for fields higher than about 11 V/mm.

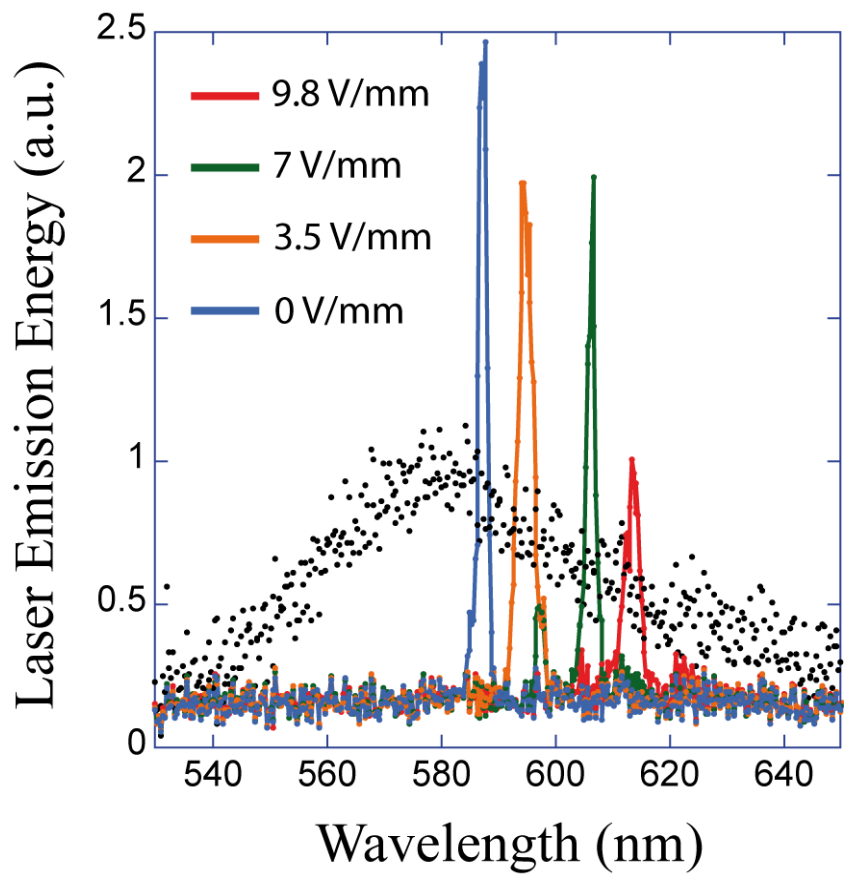

**Figure 3.** Laser emission spectra for different square-wave electric fields of frequency 1 Hz. Black points represent the fluorescence spectrum of the dye, and are drawn on a different scale.

Finally, a different mixture was prepared in order to take the 3rd harmonic $\lambda_3$ band into the fluorescence range of the dye with the purpose of investigating the possibility of lasing emission at this photonic band. We used a different left-handed chiral dopant: 4-[(4-nitrophenoxy)carbonyl]phenyl (S)-2-(sec-butoxy)-4-methoxybenzoate (S1) [37], because the results with R5011 were not satisfactory probably due to a reduction of the local birefringence induced by the chiral dopant that becomes more relevant in the strength of the $\lambda_3$ band. Contrary to R5011, S1 is a chiral nematic ferroelectric liquid crystal below 40 ºC and was used by Zhao et al. [37] in mixtures with RM734 in different proportions to promote $N_F$* phases with controlled pitch length. For our sample, we prepared a mixture of RM734, S1, PM597 in proportions 78.7:20.5:0.8 (wt. %). In this mixture the $N_F$* phase becomes stable at 116 ºC on cooling from the high temperature N* phase, and crystallization occurs at about 80 ºC. Figure 4(a) shows the texture of the sample in the $N_F$* phase as observed between crossed polarizers. As can be seen, some oily streaks are also present in similar proportion as those in the previous mixture and were also present in the N* phase. Figure 4(b) represents the reflection spectrum when a square wave electric field of frequency 10 Hz and amplitude 7 V/mm was applied to the material. Apart from the usual $\lambda = pn$ band centered at 850 nm, different $\lambda_m$ photonic bands are clearly visible when the field is applied ($\lambda_3 = 580$ nm and $\lambda_4 = 443$ nm). Figure 4(c) represents the $m = 3$ band for different electric field amplitudes. For moderate fields, it is gradually and reversibly red shifted as a consequence of the pitch enlargement when the field is increased. The tuning amplitude for this band is approximately 2/3 of that of the zero-field one ($\lambda_2 = pn$). The relation is not exact because of index dispersion. Laser emission was attempted by optically pumping the material with right-handed circularly polarized light and repetition rate of 1 Hz, synchronized with the applied square-wave electric field. Figure 4(d) shows two laser shots for a field-amplitude of 7 V/mm and a pumping energy of 8 µJ/pulse emitted from different regions of the sample. The small differences in the laser wavelength can be attributed to inhomogeneities in the sample pitch or in the electric field in the different regions of the sample. In this case, we could not further characterize this laser or evaluate its electrical tuning possibilities because lasing stops after a few pulses. We speculate that heating by such an energetic pumping beam can presumably induce changes in the $N_F$* structure and also decompose the organic dye molecules at the illuminated spot region. A total recuperation of the laser performance requires a slow process involving molecular diffusion that replace the deteriorated molecules by new ones at the spot position.

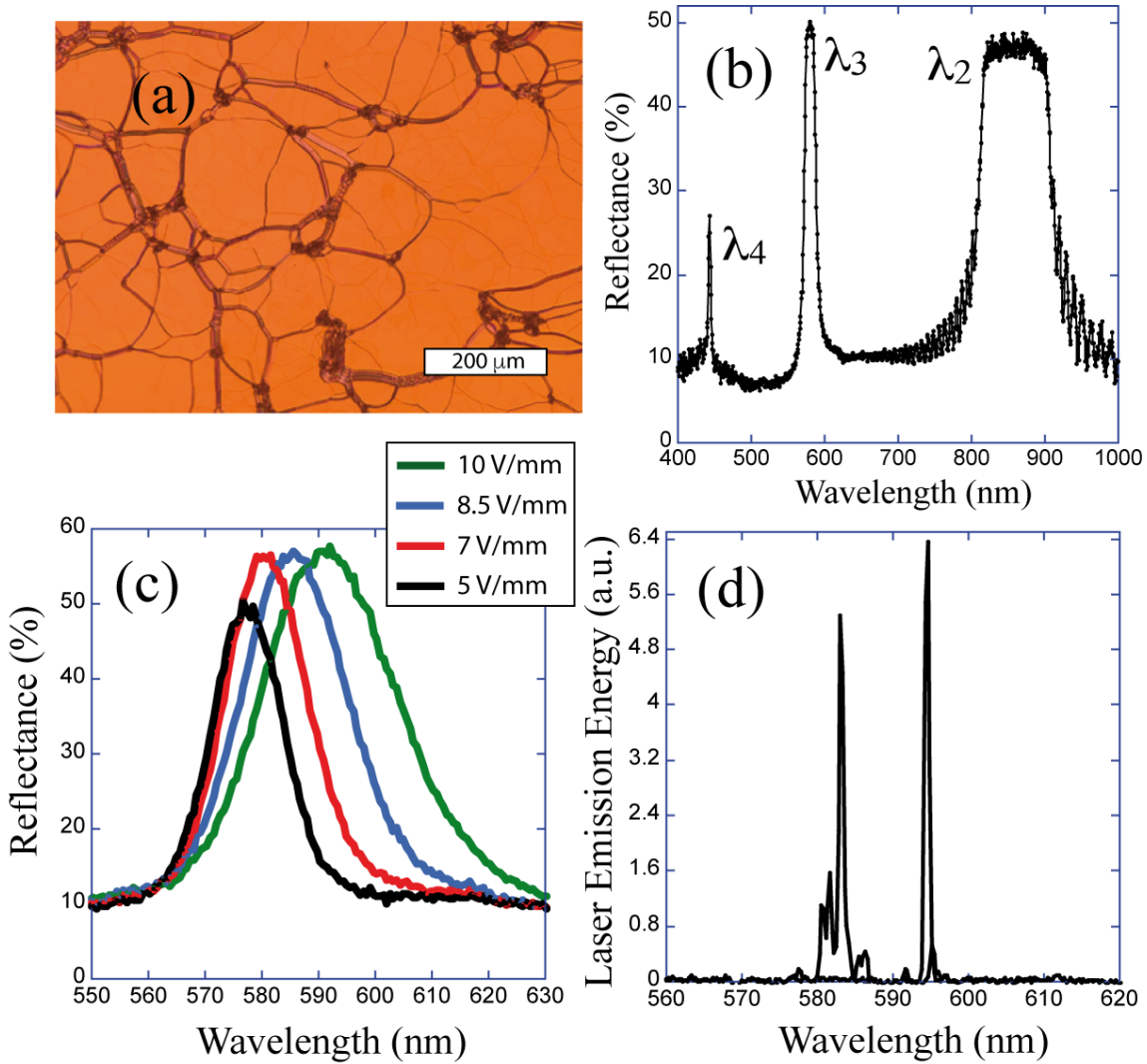

**Figure 4.** Optical characteristics and electric-field tuning of the $\lambda_3$ band. (a) Photomicrograph of the texture of the material at 104 ºC in the $N_F^*$ phase as observed between crossed polarizers. (b) Reflectance spectrum for a square-wave electric field intensity of 7 V/mm and frequency 10 Hz. The different photonic bands for $m = 2, 3,$ and 4 are indicated. (c) Reflectance spectra around the $\lambda_3$ photonic band measured for different square-wave electric field intensities of frequency 10 Hz. (d) Single-shot laser emission spectra at two different regions of the cell when the material was optically pumped with 8 μJ/pulse and repetition rate 1 Hz, synchronized with an applied square-wave electric field of 7 V/mm.

**Conclusions**

We presented some laser devices based on the chiral ferroelectric nematic phase. We have demonstrated the possibility of attaining laser emission not only at the edges of the usual $\lambda_2$ half pitch PBG but also at the third harmonic band that appears when an electric field is applied perpendicular to the helical axis, and whose wavelength depends on the field intensity. In addition, we have shown the possibility of tunable laser emission at the LWE of the half pitch PBG under electric fields. The process is fully reversible and we have attained a tuning range of about 30 nm.

In order to check the laser quality, we have characterized the usual laser emission at zero field, and have observed similar performances as CLC based standard lasers. It is interesting to point out that very low electric fields (of the order of a few V/mm) are sufficient to induce structural distortions that result in relevant changes in the photonic properties of a $N_F$* material. This greatly facilitates experiments, as well as the design of laser devices. In contrast to this, we reported previously [39] laser emission at the second order PBG of the electrically-distorted helix of a conventional CLC. In that case, electric field bursts of 20 kHz of frequency and 4500 V/mm were necessary to be applied during a short time (0.1 – 2 ms) in order to sufficiently distort, but not unwind, the helix.

We would like to point out that a substantial improvement in the performance of the lasers described here could be achieved with a better alignment of the $N_F$* phase. Good alignment enhances the quality factor of the laser cavity because reduces the scattering of the pumping light, which leads to the decrease of $E_{th}$ and to the increase of the laser slope efficiency [40]. In this regard it would be worthwhile to investigate the quality of alignment that can be achieved with other $N_F$* materials, some of which could work even at room temperature. Beside this, another interesting strategy to improve lasing is the use of multilayer cells [5,41], which, in its simplest configuration, consists of assembling three cholesteric layers of different handedness. The two external films have the same handedness as the active medium, are inactive and act as reflectors, and only the internal layer is dye-doped. Thus, the role of the reflectors is similar to that of the mirrors in Fabry-Perot cavity lasers. Using this strategy, a considerable reduction of $E_{th}$ was achieved

[5,41] in conventional CLC lasers, which could be very useful in the framework of the present work for lasing emission at the 3rd or 4th harmonic photonic bands.

As a general conclusion, this work demonstrates some interesting possibilities of laser emission based on $N_F^*$ materials. The rich variety of photonic properties exhibited by these materials makes them singular candidates to design simpler and versatile laser devices. These are some of our ideas for future work.

**Acknowledgments**

This work was financially supported by the Basque Government project IT1458-22, the Spanish project PID2021-122882NB-I00/AEI /10.13039/501100011033/ and by "ERDF A way of making Europe", the Gobierno de Aragón-FSE (E47_23R- research group). The authors would like to acknowledge the Servicios Científico-Técnicos of CEQMA (CSIC-Universidad de Zaragoza) for their support.

# Supporting Information for

# Chiral ferroelectric nematic liquid crystals as materials for versatile laser devices

César L. Folcia, Josu Ortega, Teresa Sierra, Alejandro Martínez-Bueno, and Jesús Etxebarria

Email:  cesar.folcia@ehu.es or josu.ortega@ehu.es

**Synthesis and characterization of compound S1.**

All the starting products and reagents were purchased from Merck. Solvents were purchased from Fisher Scientific and dried by using a solvent purification system. $^1$H-NMR and $^{13}$C-NMR spectra were acquired on a Bruker AV400 spectrometer. The experiments were performed at room temperature in different deuterated solvents (CDCl$_3$, CD$_2$Cl$_2$ and acetone-d6). Chemical shifts are given in ppm relative to TMS and the solvent residual peak was used as internal standard. Positive ion electrospray ionization high resolution mass spectrometry (HRMS ESI$^+$) was performed on a Bruker Q-TOF-MS. Polarimetry was performed on a Jasco P-10120 polarimeter in chloroform as solvent. Figure S1 represents the consecutive steps of the synthesis.

**Methyl (*S*)-2-(sec-butoxy)-4-methoxybenzoate (1)**: methyl 2-hydroxy-4-methoxybenzoate (13.5 mmol), (*R*)-2-butanol (13.5 mmol) and triphenylphosphine (16.2 mmol) were dissolved in dry dichloromethane under argon atmosphere. Then, DIAD (16.2 mmol) was added dropwise, and the reaction mixture was stirred overnight at room temperature. After this time, solvent was removed under reduced pressure and the crude was purified by flash chromatography using hexane/CH$_2$Cl$_2$ (8:2) and increasing the polarity to hexane/CH$_2$Cl$_2$ (6:4). The product was obtained as colourless oil. Yield: **74 %.** $^1$**H-RMN** (400 MHz, CDCl$_3$) δ 7.87 - 7.76 (m, 1H, ArH), 6.52 - 6.42 (m, 2H, ArH), 4.40 – 4.26 (m, 1H, OCH), 3.84 (s, 3H, OCH$_3$), 3.83 (s, 3H, OCH$_3$), 1.87 - 1.61 (m, 2H, CH$_2$),1.33 (d, *J*=6.1 Hz, 3H, CH$_3$), 1.00 (t, *J*=7.4 Hz, 3H, CH$_3$). $^{13}$**C-RMN** (100 MHz, CDCl$_3$) δ 166.6, 164.0, 160.2, 133.9, 114.0, 104.9, 101.8, 76.8, 55.6, 51.7, 29.3, 19.2, 9.8.

**(*S*)-2-(sec-butoxy)-4-methoxybenzoic acid (2):** Lithium hydroxide monohydrate (36.9 mmol) was added to a suspension of **compound 1** (9.2 mmol) in a mixture of methanol/water (40 mL/20 mL). The mixture reaction was stirred for 2 hours at room temperature. Then water was added (100 mL) and the crude reaction was acidified with 2M HCl until pH = 4 and extracted with ethyl acetate. Then, the organic phase was dried with MgSO$_4$, filtered and the solvent was removed under reduced pressure. The product was obtained as a colourless oil. Yield: **97 %**. **$^1$H-RMN** (400 MHz, CDCl$_3$) δ 11.00 (s, 1H, COOH), 8.19 – 8.09 (m, 1H, ArH), 6.67 - 6.59 (m, 1H, ArH), 6.53 – 6.48 (m, 1H, ArH), 4.65 – 4.53 (m, 1H, OCH), 3.87 (s, 3H, OCH$_3$), 1.94 – 1.71(m, 2H, CH$_2$),1.43 (d, *J*=6.1 Hz, 3H, CH$_3$), 1.03 (t, *J*=7.5 Hz, 3H, CH$_3$). **$^{13}$C-RMN** (100 MHz, CDCl$_3$) δ 165.6, 165, 158.1, 135.6, 111.5, 106.8, 100.8, 78.9, 55.8, 29.1, 19.3, 9.7.

**Compound 3**: **Compound 2** (6.7 mmol), benzyl 4-hyroxybenzoate (6.7 mmol) and DMAP (2.6 mmol) were dissolved in dry dichloromethane under argon atmosphere. Then, a solution of DCC (10 mmol) in dry dichloromethane was added dropwise and the reaction mixture was stirred overnight at room temperature. After this time, the solid formed during the reaction was filtered off, solvent was removed under reduced pressure and the crude was purified by flash chromatography using CH$_2$Cl$_2$/hexane (8:2) and increasing the polarity to hexane/CH$_2$Cl$_2$ (9:1). The product was obtained as withe powder. Yield: **74 %.** **$^1$H-RMN** (400 MHz, CD$_2$Cl$_2$) δ 8.16 – 8.10 (m, 2H, ArH), 8.01 – 7.95 (m, 1H, ArH), 7.52 - 7.32 (m, 5H, ArH), 7.32 – 7.25 (m, 2H, ArH), 6.59 - 6.51 (m, 2H, ArH), 5.37 (s, 2H, CH$_2$), 4.49 – 4.35 (m, 1H, OCH), 3.86 (s, 3H, OCH$_3$), 1.85 - 1.61 (m, 2H, CH$_2$),1.34 (d, *J*=6.1 Hz, 3H, CH$_3$), 0.99 (t, *J*=7.4 Hz, 3H, CH$_3$). **$^{13}$C-RMN** (100 MHz, CD$_2$Cl$_2$) δ 166.2, 165.5, 164.0, 161.4, 155.8, 136.9, 134.9, 131.6, 129.1, 128.7, 128.6, 127.9, 122.6, 112.6, 105.6, 101.6, 77.0, 67.2, 56.1, 29.8, 19.4, 9.9.

**Compound 4**: Pd/C 10% (0.2 g) was added to a solution of **compound 3** (4.6 mmol) in ethyl acetate. The mixture was stirred at room temperature for 4 hours under H$_2$ atmosphere. Then the catalyst was filtered off through a celite pad and the solvent was removed under reduced pressure. The product was obtained as withe powder. Yield: **94 %.** **$^1$H-RMN** (400 MHz, CD$_2$Cl$_2$) δ 8.20 – 8.13 (m, 2H, ArH), 8.05 – 8.00 (m, 1H, ArH), 7.34 – 7.28 (m, 2H, ArH), 6.59 - 6.48 (m, 2H, ArH), 4.46 – 4.35 (m, 1H, OCH), 3.87 (s, 3H, OCH$_3$), 1.89 - 1.62 (m, 2H, CH$_2$),1.36 (d, *J*=6.1 Hz, 3H, CH$_3$), 1.00 (t,

*J*=7.4 Hz, 3H, CH$_3$). **$^{13}$C-RMN** (100 MHz, CD$_2$Cl$_2$) δ 171.3, 165.0, 163.6, 161.1, 155.8, 134.7, 131.9, 122.2, 112.1, 105.0, 101.3, 76.7, 55.7, 29.3, 19.2, 9.8.

**Compound S1**: **Compound 4** (4.6 mmol), *p*-nitrophenol (4.6 mmol) and DMAP (1.8 mmol) were dissolved in dry dichloromethane under argon atmosphere. Then, a solution of DCC (6.9 mmol) in dry dichloromethane was added dropwise and the reaction mixture was stirred overnight at room temperature. After this time, the solid formed during the reaction was filtered off, solvent was removed under reduced pressure and the crude was purified by flash chromatography using CH$_2$Cl$_2$/hexane (8:2) and increasing the polarity to CH$_2$Cl$_2$. The product was obtained as viscous oil. Yield: **25 %**. **$^1$H-RMN** (400 MHz, acetone-d$_6$) δ 8.44 – 8.36 (m, 2H, ArH), 8.33 – 8.26 (m, 2H, ArH), 8.01 – 7.94 (m, 1H, ArH), 7.70 – 7.62 (m, 2H, ArH), 7.53 – 7.44 (m, 2H, ArH), 6.75 - 6.69 (m, 1H, ArH), 6.69 - 6.62 (m, 1H, ArH), 4.67 – 4.55 (m, 1H, OCH), 3.91 (s, 3H, OCH$_3$), 1.85 - 1.64 (m, 2H, CH$_2$), 1.35 (d, *J* = 6.0 Hz, 3H, CH$_3$), 1.01 (t, *J*=7.4 Hz, 3H, CH$_3$) (see Figure S2). **$^{13}$C-RMN** (100 MHz, acetone-d$_6$) δ 166.2, 164.4, 164.1, 161.8, 157.3, 157.1, 146.6, 135.2, 132.7, 126.9, 126.2, 124.2, 123.6, 112.9, 106.6, 101.8, 76.9, 56.2, 30.1, 19.5, 10.0, (Figure S2). **HRMS-ESI$^+$**: 466.1498 [M+H]$^+$, 410.0873 [M-C$_4$H$_7$]$^+$. $[\alpha]_D^{25}$ (CH$_3$Cl) = +19.8232 deg, (see Figure S3).

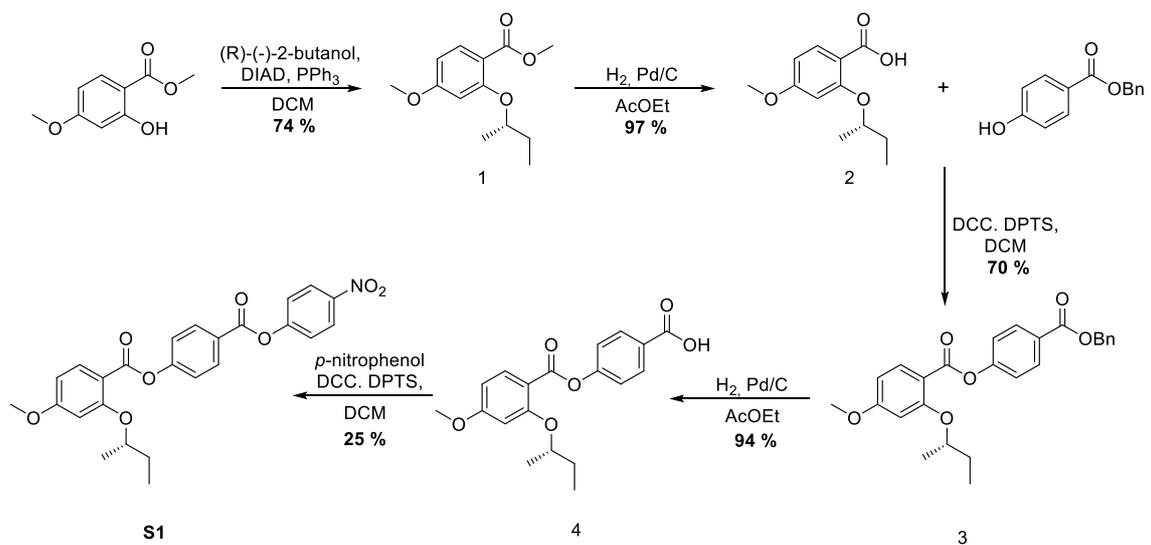

**Fig. S1.** Synthesis of compound S1.

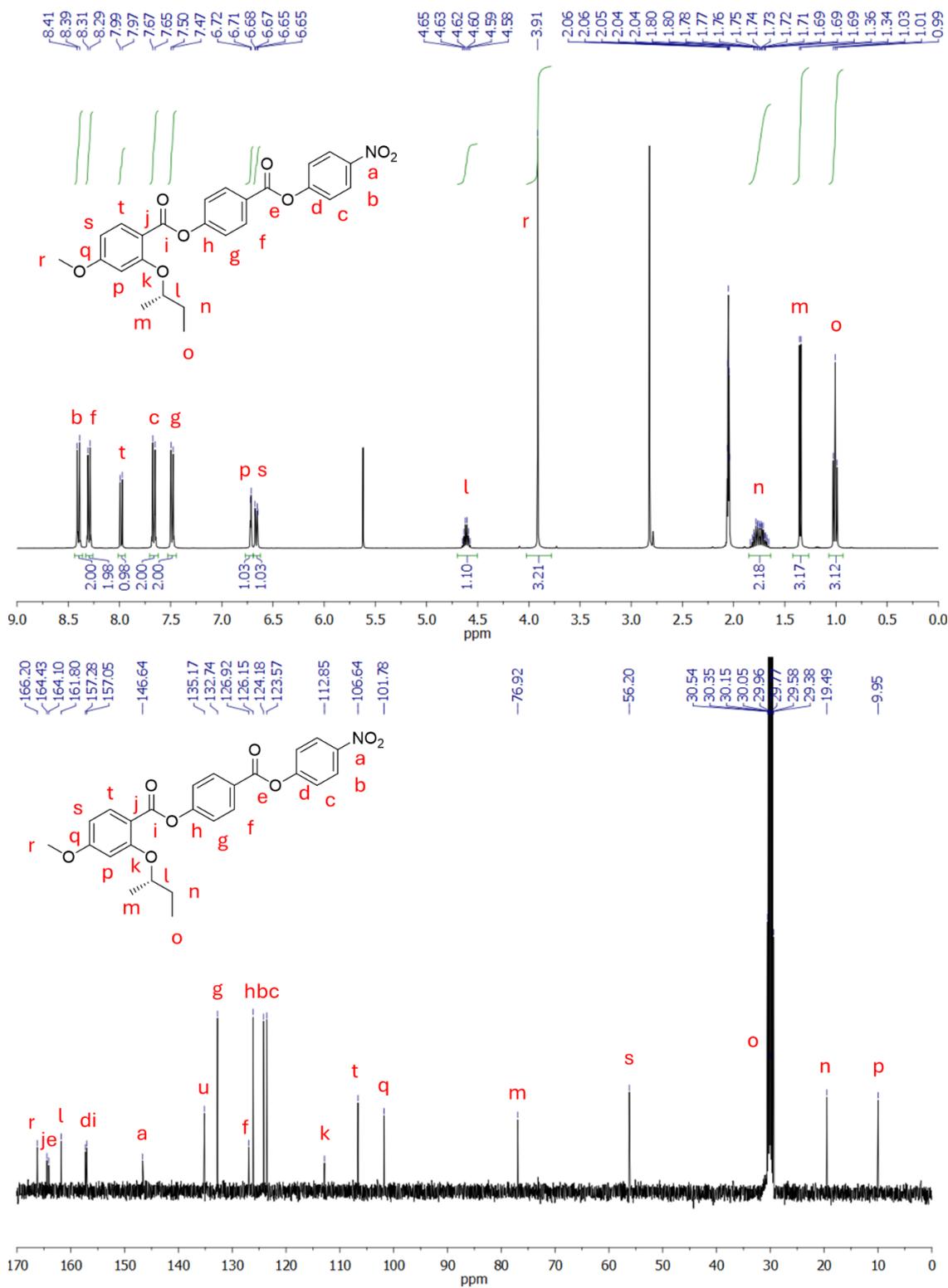

**Fig. S2.** $^1$H NMR and $^{13}$C NMR spectra of S1 in acetone-$d_6$.

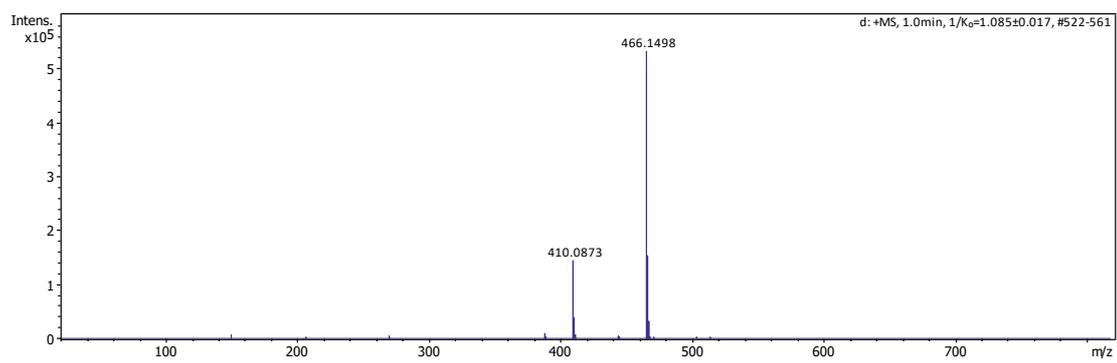

**Fig. S3.** HRMS-ESI+ spectrum of S1.